\makeatletter
\newcommand{\dontusepackage}[2][]{%
  \@namedef{ver@#2.sty}{9999/12/31}%
  \@namedef{opt@#2.sty}{#1}}
\makeatother
\dontusepackage{subfigure}

\documentclass[]{article}

\usepackage{lmodern}
\usepackage{amssymb,amsmath}
\usepackage{ifxetex,ifluatex}
\usepackage[usenames,dvipsnames]{color}
\usepackage{fixltx2e} 
\ifnum 0\ifxetex 1\fi\ifluatex 1\fi=0 
  \usepackage[T1]{fontenc}
  \usepackage[utf8]{inputenc}
\else 
  \ifxetex
    \usepackage{mathspec}
    \usepackage{xltxtra,xunicode}
  \else
    \usepackage{fontspec}
  \fi
  \defaultfontfeatures{Mapping=tex-text,Scale=MatchLowercase}
  
\fi
\IfFileExists{upquote.sty}{\usepackage{upquote}}{}
\IfFileExists{microtype.sty}{%
\usepackage{microtype}
\UseMicrotypeSet[protrusion]{basicmath} 
}{}
\usepackage[margin=1.0in,bottom=1.5in]{geometry}
\usepackage[]{natbib}
\usepackage{listings}
\usepackage{graphicx}
\makeatletter
\def\maxwidth{\ifdim\Gin@nat@width>\linewidth\linewidth\else\Gin@nat@width\fi}
\def\maxheight{\ifdim\Gin@nat@height>\textheight\textheight\else\Gin@nat@height\fi}
\makeatother
\setkeys{Gin}{width=\maxwidth,height=\maxheight,keepaspectratio}
\usepackage{caption}
\usepackage{float}
\ifxetex
  \usepackage[setpagesize=false, 
              unicode=false, 
              xetex]{hyperref}
\else
  \usepackage[unicode=true]{hyperref}
\fi
\hypersetup{breaklinks=true,
            bookmarks=true,
            pdfauthor={},
            pdftitle={Compressive time-lapse seismic monitoring of carbon storage and sequestration with the joint recovery model},
            colorlinks=true,
            citecolor=black,
            urlcolor=blue,
            linkcolor=black,
            pdfborder={0 0 0}}
\title{Compressive time-lapse seismic monitoring of carbon storage and
sequestration with the joint recovery model}
\author{Ziyi Yin\textsuperscript{1}, Mathias Louboutin\textsuperscript{2}, Felix
J. Herrmann\textsuperscript{1,2}\\\textsuperscript{1} School of
Computational Science and Engineering, Georgia Institute of
Technology\\\textsuperscript{2} School of Earth and Atmospheric
Sciences, Georgia Institute of Technology\\}
\date{}

\begin{document}
\maketitle

\section{Summary}\label{summary}

\vspace*{-0.3cm}

Time-lapse seismic monitoring of carbon storage and sequestration is
often challenging because the time-lapse signature of the growth of CO2
plumes is weak in amplitude and therefore difficult to detect
seismically. This situation is compounded by the fact that the surveys
are often coarsely sampled and not replicated to reduce costs. As a
result, images obtained for different vintages (baseline and monitor
surveys) often contain artifacts that may be attributed wrongly to
time-lapse changes. To address these issues, we propose to invert the
baseline and monitor surveys jointly. By using the joint recovery model,
we exploit information shared between multiple time-lapse surveys.
Contrary to other time-lapse methods, our approach does not rely on
replicating the surveys to detect time-lapse changes. To illustrate this
advantage, we present a numerical sensitivity study where CO2 is
injected in a realistic synthetic model. This model is representative of
the geology in the southeast of the North Sea, an area currently
considered for carbon sequestration. Our example demonstrates that the
joint recovery model improves the quality of time-lapse images allowing
us to monitor the CO2 plume seismically.

\vspace*{-0.45cm}

\section{Introduction}\label{introduction}

\vspace*{-0.3cm}

Time-lapse seismic imaging technology has recently been deployed by
Carbon Capture and Storage (CCS) practitioners to monitor CO2 dynamics
in porous media over long periods of time, with the Sleipner project in
Norway as a pioneer \citep{ARTS2003347, FURRE20173916}. There have been
several major developments to adapt compressive sensing
\citep{janiszewski2014improvements, herrmann2008GJInps} and machine
learning \citep{bharadwaj2020symae, kaur2020time} to improve the
productivity of (time-lapse) seismic data acquisition
\citep{oghenekohwo2016GEOPctl, wason2016GEOPctl} and imaging
\citep{oghenekohwo2015CSEGctl} by exploiting similarities between
baseline and monitor surveys. Recent work by \citet{li2020coupled}
indicates the potential benefits of a coupled-physics framework where
the slow-time subsurface CO2 flow, a rock physics model, and the
fast-time wave physics are combined to better understand the dynamics of
CO2 plumes and their effect on time-lapse seismic data for the duration
of a CCS project. Because the time-lapse signal of CO2 plumes is weak,
it continues to pose challenges on the robustness of time-lapse imaging
\citep{lumley1997assessing}.

To address some of these challenges,
\citet{oghenekohwo2016GEOPctl};\citet{wason2016GEOPctl} proposed the
joint recovery model (JRM) designed to reap the benefits of low-cost
randomized non-replicated acquisition. It takes advantage of the fact
that time-lapse seismic data and subsurface structure undergoing
localized changes share information among the different vintages. In
this extended abstract, we present a synthetic case study to make the
case that the dynamics of CO2 plumes can indeed be monitored with
active-source surface seismic. The presented work can be considered as a
continuation of recent contributions by \citet{li2020coupled}, who
considered seismic monitoring in a much simpler geological setting
involving cross-well tomography. In our study, we consider a 2D
subsection of the BG Compass model with a geology representative of the
southeast of the North Sea. Blunt sandstones in this area are considered
as possible reservoirs for injection of CO2. Aside from assessing the
sensitivity of seismic imaging towards monitoring CO2 plumes, we also
study the effects of changes in the acquisition and the presence of
noise, and how the JRM can be deployed to mitigate these effects of
incomplete and non-replicated data.

Our contributions are organized as follows. First, we briefly discuss
full-waveform inversion and its linearization as part of time-lapse
seismic imaging. Next, we introduce our joint inversion framework that
explores commonalities between seismic images associated with the
different vintages collected for the duration of a CCS project. We
conclude by discussing a realistic time-lapse imaging experiment
juxtaposing images obtained by inverting the vintages independently or
jointly with JRM.

\vspace*{-0.45cm}

\section{Methodology}\label{methodology}

\vspace*{-0.3cm}

We first briefly introduce the theoretical framework on which our
wave-equation based monitoring framework for CCS is based. While we
understand that more complex wave physics, e.g.~elastic
\citep{thomsen1986weak} and nonlinear inversions \citep{Li11TRfrfwi},
may eventually be needed, we begin by deriving a model based on the
scalar wave equation parameterized by the squared slowness. We start by
introducing the objective for full-waveform inversion and its
linearization with respect to an assumed given smooth background model.
Based on this linearization, we propose a joint recovery model, which we
solve computationally efficiently with linearized Bregman iterations
\citep{yin2008bregman, witte2018cls}.

\vspace*{-0.15cm}

\subsection{Full-waveform inversion (FWI) and its
linearization}\label{full-waveform-inversion-fwi-and-its-linearization}

\vspace*{-0.15cm}

To derive a joint wave-based seismic monitoring system, we start by
introducing the objective of full-waveform inversion for the baseline
($j=1$) and multiple monitor surveys ($2 \leq j \leq n_v$) with
$n_v\geq 2$ as the number of vintages. For each vintage, this data
misfit objective reads
\begin{equation}
\underset{\mathbf{m}_j}{\operatorname{min}}\quad \|\mathbf{d}_j-\mathcal{F}_j(\mathbf{m}_j)\|_2^2 \quad \text{for}\quad j=\{1,2,\cdots,n_v\}.
\label{eqfwi}
\end{equation}
 In this expression, $\mathcal{F}_j$ is the nonlinear forward modeling
operator, which for given sources generates synthetic data at the
receiver locations from the discretized time-lapse model parameters
$\mathbf{m}_j$. The observed data is, for each vintage, collected in the
vector $\mathbf{d}_j$. When provided with background velocity models
$\mathbf{\overline{m}}_j$ for each vintage, the above forward model can
be linearized yielding
\begin{equation}
\underset{\mathbf{\delta m}_j}{\operatorname{min}} \quad \|\mathbf{\delta d}_j-\nabla \mathcal{F}_j(\mathbf{\overline{m}}_j)\mathbf{\delta m}_j\|_2^2,
\label{eqgaussnewton}
\end{equation}
 where
$\mathbf{\delta d}_j=\mathbf{d}_j-\mathcal{F}_j(\mathbf{\overline{m}}_j)$
are the linearized datasets for each vintage and $\nabla \mathcal{F}_j$
the linearized forward operator relating perturbations in the squared
slowness for each vintage, $\mathbf{\delta m}_j$, to linearized data
$\mathbf{\delta d}_j$.

\vspace*{-0.15cm}

\subsection{Time-lapse monitoring with the joint recovery
model}\label{time-lapse-monitoring-with-the-joint-recovery-model}

\vspace*{-0.15cm}

There exists an extensive literature on wave-based time-lapse imaging
aimed at producing images that show time-lapse differences in the image
space by carrying out inversion rather than imaging
\citep{qu2017simultaneous, yang2016time, queisser2013full, maharramov2019integrated}.
A prominent example of such inversion method is formed by time-lapse
monitoring via double differences
\citep{zhang2013double, yang2015double} where differences are obtained
by inverting differences between the monitor and baseline residuals
rather than between observed and synthetic monitoring data. While this
type of approach can lead to good results, it assumes the baseline and
monitor surveys to be relatively noise free, well sampled, and above all
replicated---i.e.~the acquisition geometries between different surveys
need to be identical.

In our approach, we build on a low-cost formulation for time-lapse
seismic monitoring designed to handle low-cost non-replicated sparsely
sampled surveys. The basic idea of this approach is to focus on what is
shared amongst the different vintages rather than focusing on what is
different. We argue that time-lapse monitoring benefits from such an
approach if the Earth model being monitored undergoes relatively
localized changes, an assumption that likely holds during CCS. To arrive
at a successful joint inversion scheme, we augment the block diagonal
linear system undergirding independent imaging (by minimizing
Equation~\ref{eqgaussnewton} independently for each vintage) with an
extra column and a corresponding unknown common component. With these
steps, we write
\begin{equation}
\begin{aligned}
\mathbf{A} = \begin{bmatrix}  \frac{1}{\gamma}\nabla\mathcal{F}_1(\mathbf{\overline{m}}_1) & \nabla\mathcal{F}_1(\mathbf{\overline{m}}_1) &  \mathbf{0} &\mathbf{0} \\ \cdots & \mathbf{0} &  \cdots & \mathbf{0} \\
\frac{1}{\gamma}\nabla\mathcal{F}_{n_v}(\mathbf{\overline{m}}_{n_v}) & \mathbf{0} & \mathbf{0} & \nabla\mathcal{F}_{n_v}(\mathbf{\overline{m}}_{n_v})
\end{bmatrix}.
\end{aligned}
\label{eqjrm}
\end{equation}
 This matrix $\mathbf{A}$ relates the linearized data for each vintage
$\{\mathbf{\delta d}_j\}_{j=1}^{n_v}$, collected in the vector
$\mathbf{b} = \left[\mathbf{\delta d}_1^\top, \cdots, \mathbf{\delta d}_{n_v}^\top\right]^\top$,
to the common component $\mathbf{z}_0$ and innovations with respect to
this common component $\{\mathbf{z}_j\}_{i=1}^{n_v}$ all collected in
the vector
$\mathbf{z} = \left[\mathbf{z}_0^\top,\cdots,\mathbf{z}_{n_v}^\top\right]^\top$.
This formulation emphasizes what is common amongst the vintages when
$\gamma$ is close to $0$, given that $0<\gamma<n_v$ is generally a good
choice in practice \citep{li2015weighted}.

In JRM, surveys for all vintages are explicitly related to the common
component. Therefore, when acquisition geometries for different surveys
are not replicated, this joint formulation recovers the common component
better because the different surveys collect complementary information
through the different acquisitions. As a result of the improved recovery
of the common component, we can also expect the images for the different
vintages themselves to be better recovered. This behavior has indeed
been confirmed by several numerical experiments
\citep{oghenekohwo2017EAGEitl, wason2015EAGEcsm, oghenekohwo2015CSEGctl}
and in the presence of noise as reported by Chevron
\citep{wei2018improve, tian2018joint}.

\vspace*{-0.15cm}

\subsection{Time-lapse monitoring with linearized
Bregman}\label{time-lapse-monitoring-with-linearized-bregman}

\vspace*{-0.15cm}

The above joint recovery model reaps the benefit of randomized seismic
acquisition with techniques adapted from Compressive Sensing, where
subsampling related interferences (e.g.~aliases or simultaneous source
cross talk) are rendered into incoherent noise
\citep{herrmann2008GJInps, herrmann2010GEOPrsg}. As shown by
\citet{yang2020tdsp};\citet{witte2018cls}, these incoherent artifacts
can be mapped back to coherent energy by solving the following
minimization problem:
\begin{equation}
\begin{split}
\underset{\mathbf{x}}{\operatorname{min}} \quad \lambda \|\mathbf{C}\mathbf{x}\|_1+\frac{1}{2}\|\mathbf{C}\mathbf{x}\|_2^2 \\
\text{subject to}\quad \|\mathbf{b}- \mathbf{A}\mathbf{x}\|_2^2 \leq \sigma
\end{split}
\label{elastic}
\end{equation}
 with $\mathbf{C}$ the forward curvelet transform, $\lambda$ a threshold
parameter, and $\sigma$ the magnitude of the noise. The advantage of
this strictly convex formulation is that it can be solved via linearized
Bregman iterations that for $\sigma=0$ correspond to doing iterative
soft thresholding on the dual variable---i.e., we carry out the
following update at iteration $k$
\begin{equation}
\begin{aligned}
\begin{array}{lcl} 
  \mathbf{u}_{k+1} & = & \mathbf{u}_k-t_k \mathbf{A}_k^\top(\mathbf{A}_k\mathbf{x}_{k}-\mathbf{b}_k)\\
 \mathbf{x}_{k+1} & = & \mathbf{C}^\top S_{\lambda}(\mathbf{C}\mathbf{u}_{k+1}),
\end{array}
\end{aligned}
\label{LBk}
\end{equation}
 where $\mathbf{A}_k$ represents the matrix in Equation~\ref{eqjrm} for
a subset of shots randomly selected from sources in each vintage. The
vector $\mathbf{b}_k$ contains the extracted shot records from
$\mathbf{b}$ and the symbol $^\top$ refers to the adjoint. Sparsity is
promoted via curvelet-domain soft thresholding
$S_{\lambda}(\cdot)=\max(|\cdot|-\lambda,0)\operatorname{sign}(\cdot)$
with $\lambda$ as the threshold.

Compared to other $\ell_1$-norm solvers, linearized Bregman iterations
are relatively simple to implement and have shown to work well in the
context of least-squares imaging \citep{yang2020tdsp, witte2018cls} as
long as the steplength $t_k$ and threshold $\lambda$ are appropriately
chosen. A general choice of dynamic steplength $t_k$ is given by
$t_k=\|\mathbf{A}_k\mathbf{x}_k-\mathbf{b}_k\|_2^2 / \|\mathbf{A}_k^\top(\mathbf{A}_k\mathbf{x}_k-\mathbf{b}_k)\|_2^2$\citep{lorenz2014linearized}
and the threshold $\lambda$ is typically chosen to be proportional to
the maximum of $|\mathbf{u}_k|$ in the first iteration ($k=1$). After
solving problem~\ref{elastic} by the iterative process~\ref{LBk},
estimates for the baseline and monitor images are obtained via
$\widehat{\mathbf{\delta m}}_j=\widehat{\mathbf{x}}_0+\widehat{\mathbf{x}}_j,\, j=1,2,\cdots,n_v$
where $\widehat{\mathbf{x}}$ minimizes problem~\ref{elastic}.

\vspace*{-0.45cm}

\section{Numerical Case Study Blunt Sandstone SW North
Sea}\label{numerical-case-study-blunt-sandstone-sw-north-sea}

\vspace*{-0.3cm}

By means of a realistic synthetic experiment, we demonstrate deployment
of the joint recovery model to monitor CCS. To create a realistic
monitoring scenario, we start by modeling the development of a CO2 plume
with a numerical scheme that solves the two-phase flow equation. Next,
we translate the CO2 concentration to time-lapse changes in acoustic
wavespeed perturbations, which we use to generate seismic data for the
vintages. These datasets, collected with different acquisitions, serve
as input to our imaging scheme.

\vspace*{-0.15cm}

\subsection{Geologic and rock physical
setting}\label{geologic-and-rock-physical-setting}

\vspace*{-0.15cm}

\begin{figure}
\centering
\includegraphics[width=0.800\hsize]{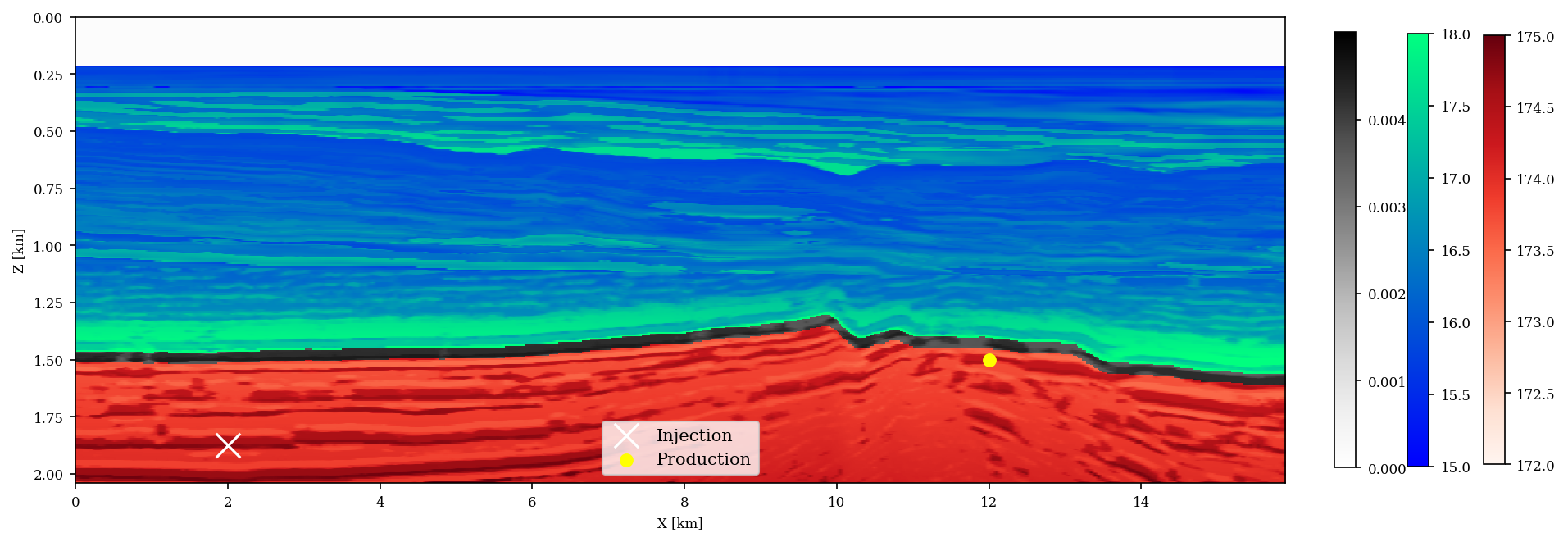}
\caption{Permeability of 2D slice of BG Compass model including
injection (white $\times$) and production well (yellow $\bullet$).
}\label{fig:perm}
\end{figure}

To numerically validate the potential improvement of time-lapse
monitoring via our joint recovery model, we consider the 2D
$2\times 15.9$km subset of the BG Compass model plotted in
Figure~\ref{fig:perm}. This figure contains a plot of the spatial
distribution for the permeability and is representative for CCS in Blunt
sandstones. Following the stratigraphy in the real setting, the
synthetic stratigraphic section in Figure~\ref{fig:perm} can roughly be
divided into three main sections, namely \emph{(i)} the highly porous
(average $22\%$) and permeable ($> 170$mD) Bunter Sandstone of about
$300-500$m thick and that serves as the CO2 reservoir (red area in
Figure~\ref{fig:perm}, which includes location of the injection well,
denoted by the white $\times$, and the production well denoted by the
yellow $\bullet$); \emph{(ii)} the primary seal (permeability
$10^{-4}-10^{-2}$mD) made of the Rot Halite Member, which is $50$m thick
and is typically continuous (black layer in Figure~\ref{fig:perm});
\emph{(iii)} the secondary seal made of the Haisborough group, which is
$>300$m thick and consists of low-permeable (permeability $15-18$mD)
mudstones.

By assuming a linear relationship between the compressional wavespeed
and permeability in each stratigraphic section, we arrive at the
synthetic permeability model plotted in Figure~\ref{fig:perm}. To
convert wavespeed to permeability, we assume for each section that an
increase of $1$km/s in the compressional wavespeed corresponds to an
increase of $1.63$mD in the permeability. Given these values for the
permeability, we derive values for the porosity using the Kozeny-Carman
equation \citep{costa2006permeability}
$K = \mathbf{\phi}^3 \left(\frac{1.527}{0.0314*(1-\mathbf{\phi})}\right)^2$,
where $K$ and $\phi$ denote permeability (mD) and porosity (\%) with
constants taken from the Strategic UK CCS Storage Appraisal Project
report.

\vspace*{-0.15cm}

\subsection{Simulation of CO2
dynamics}\label{simulation-of-co2-dynamics}

\vspace*{-0.15cm}

\begin{figure}
\centering
\includegraphics[width=1.000\hsize]{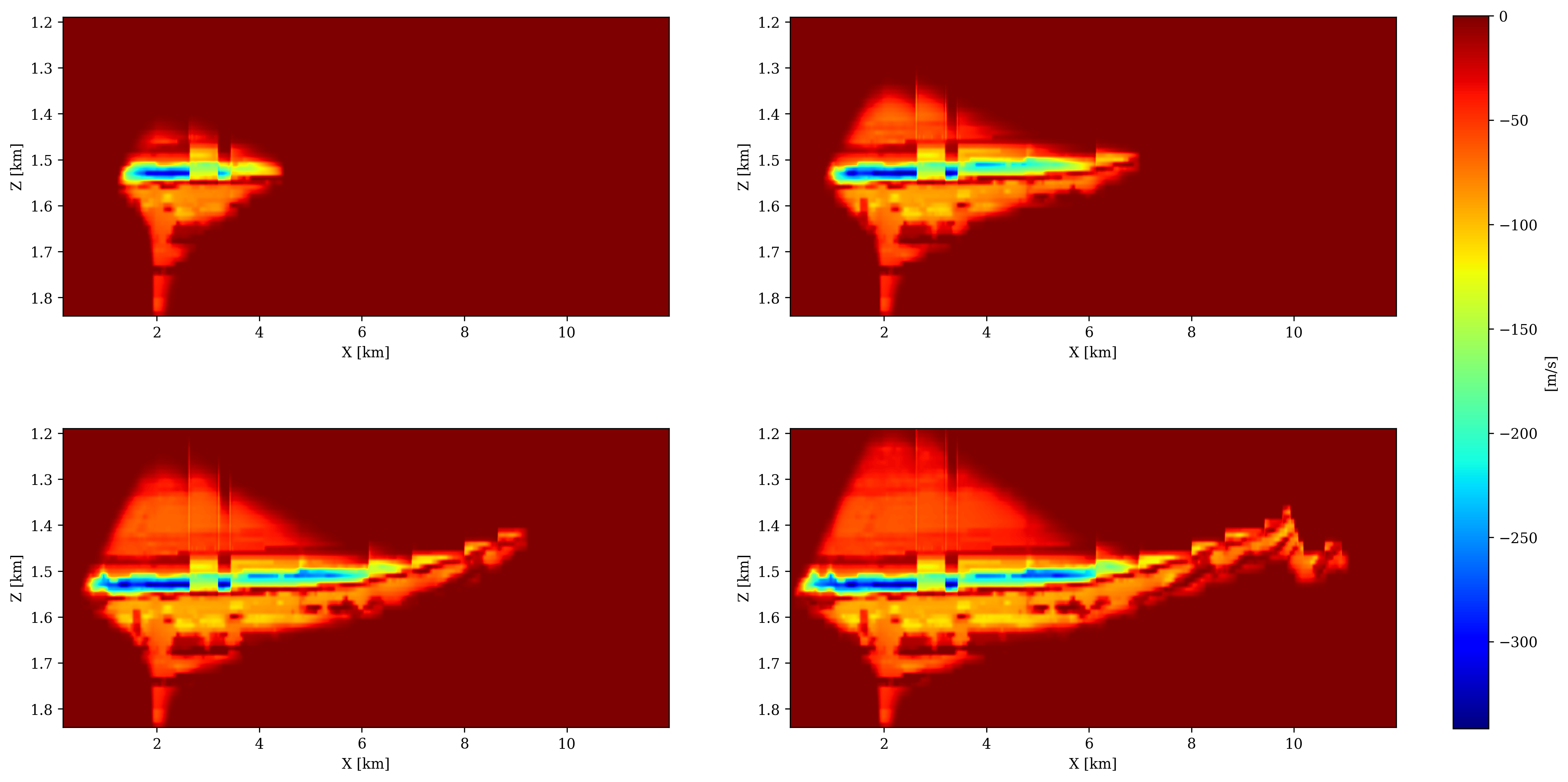}
\caption{Simulated time-lapse compressional velocity decrease after
$\{15,30,45,60\}$ years of CO2 injection.}\label{fig:plume}
\end{figure}

We use the open-source software
\href{https://github.com/lidongzh/FwiFlow.jl}{FwiFlow}
\citep{li2020coupled} to numerically simulate the growth of the CO2
plume by solving the partial differential equations for two-phase flow
for a period of $60$ years with a time step of $20$ days and a grid
spacing of $25$m. While our numerical simulations are in 2D, we assume
the permeability model to extend by $1.6$ km in the third
(perpendicular) direction. The reservoir is initially filled with saline
water, with a density of $1.053\mathrm{g/cm}^3$ and a viscosity of $1.0$
centipoise (cP). We inject CO2 at a constant rate of $7$ Mt/y for $60$
years, amounting to a total of $420$ million metric tons. During the
injection, we assume the supercritical CO2 to have a density of
$776.6 \mathrm{g/cm}^3$ and a viscosity of $0.1$cP suggested by
\citep{li2020coupled} and the Strategic UK CCS Storage Appraisal
Project. This two-phase flow simulation provides us with a map of the
CO2 concentration (in percentage) at the aformentioned time steps. We
adopt the patchy saturation model
\citep{avseth2010quantitative, li2020coupled} to convert these CO2
concentrations to decreases in bulk moduli and then to decreases in
compressional wavespeeds, which are included in Figure~\ref{fig:plume}.
From this figure, we can see how the plume develops over time under
influence of buoyancy and the production well to the right. We also
observe that the plume gives rise to a difference in p-wavespeed between
$50$m/s to $300$m/s, which should in principle be detectable via seismic
monitoring.

\vspace*{-0.15cm}

\subsection{Time-lapse data
acquisition}\label{time-lapse-data-acquisition}

\vspace*{-0.15cm}

We collect five surveys over a period of $60$ years, with a baseline
survey conducted before CO2 injection starts, followed by $4$ monitor
surveys taken after $\{15,30,45,60\}$ years of injection. We propose a
low-cost sparse non-replicated acquisition scheme with compressive
sensing, capable of producing time-lapse seismic data with a high degree
of repeatability without relying on replicating the surveys. To keep the
costs down, we work with relatively sparsely sampled ocean bottom
hydrophones connected to underwater buoys located above the ocean
bottom. Compared to ocean bottom nodes, this type of acquisition avoids
picking up coherent noise related to interface waves while it shares the
advantage of being static and therefore highly replicable. To avoid
aliasing, we jitter sample \citep{herrmann2008GJInps} the receiver
positions of $64$ hydrophones located at a depth of $120$m with an
average spacing of $250$m horizontally. To reduce the costs on the
source side, we propose non-replicated continuous simultaneous source
shooting, which after deblending \citep{li2013joint, xuan2020deblending}
yields $1272$ sources with a dense source sampling of $12.5$m
horizontally. To mimic the challenges related to active-source
acquisition, we vary for each survey the tow-depth of sources uniformly
randomly from $5$m to $15$m. We also apply random horizontal shifts
varying uniformly in a range of $-6$m to $6$m, between sources in the
different surveys. Since the receivers are coarsely sampled, we employ
source-receiver reciprocity during imaging to further reduce
computational costs.

\vspace*{-0.15cm}

\subsection{Time-lapse monitoring with the joint recovery
model}\label{time-lapse-monitoring-with-the-joint-recovery-model-1}

\vspace*{-0.15cm}

As stated earlier, we assume to have access to kinematically correct
background velocity models $\{\mathbf{\overline{m}}\}_{j=1}^n$. To
demonstrate what is in principle achievable under ideal wave-physics, we
generate linear data for each vintage via
$\mathbf{\delta d}_j=\nabla\mathcal{F}_j(\mathbf{\overline{m}_j}),\, j=1\cdots n_v$.
We implemented both inversion schemes with
\href{https://github.com/slimgroup/JUDI.jl}{JUDI} -- the Julia Devito
Inversion framework \citep{witte2018alf}. This open-source package uses
the highly optimized time-domain finite-difference propagators of
\href{https://github.com/devitocodes/devito}{Devito}
\citep{luporini2018aap, louboutin2018dae}. We use a grid spacing of
$10$m during the wave-equation simulations and a Ricker wavelet with a
peak frequency of $25$Hz. Despite committing the inversion crime by
generating data with the demigration operator, our monitoring problem is
complicated by a rather complex but realistic geology (shown in
Figure~\ref{fig:perm}) and by the fact that the sampling is coarse on
the receiver site and not replicated at the source site. To increase
realism, we also use a random trace estimation technique during imaging.
During random trace estimation, memory use is reduced drastically, a
prerequisite for scaling our approach to $3$D. Using the open-source
software
\href{https://github.com/slimgroup/TimeProbeSeismic.jl}{TimeProbeSeismic.jl},
we chose $32$ random probing vectors to approximate our gradient
calculations, which leads to noisy imaging artifacts. We refer to
\citet{louboutin2021ultra} and to the proceedings of this conference for
further details.

\vspace*{-0.15cm}

\subsection{Independent recovery}\label{independent-recovery}

\vspace*{-0.15cm}

\begin{figure}
\centering
\includegraphics[width=1.000\hsize]{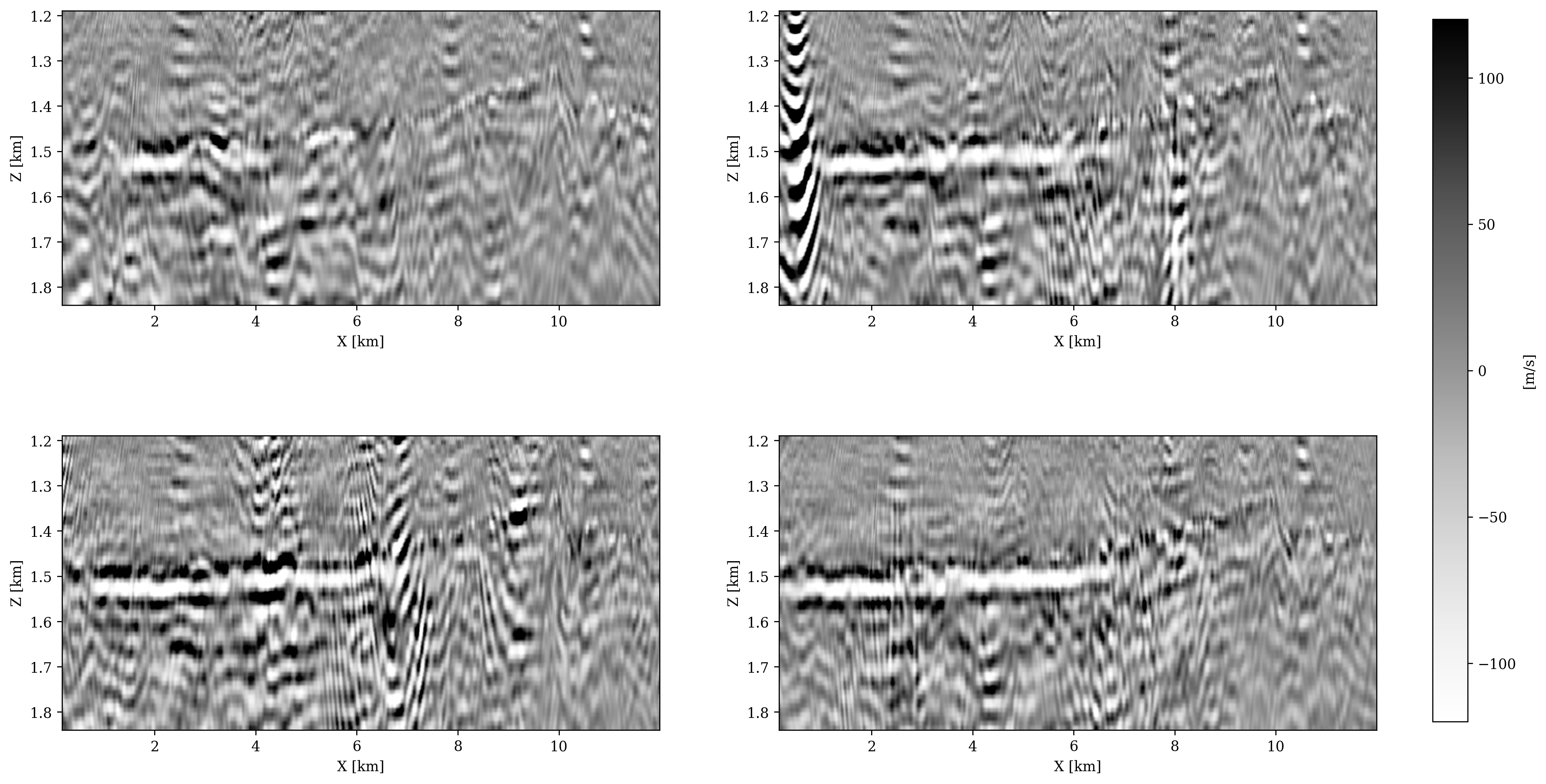}
\caption{Difference plots w.r.t. the baseline survey yielded by
independent recovery. The NRMS values computed for each pair are
$\{18.38\%, 15.74\%, 26.63\%, 16.55\%\}$.}\label{fig:LSRTMdiff}
\end{figure}

We first recover the velocity perturbations in each vintage
independently. During each Bregman iteration, we randomly select eight
shots without replication during imaging, while we select $16$ during
the first iteration to build up a good starting image. We use the
suggested steplength of \citet{lorenz2014linearized} and the $90$th
percentile of $|\mathbf{u}_k|$ at the first iteration as the threshold
$\lambda$. For each vintage, we do $23$ Bregman iterations, which
amounts to three data passes. To observe the growth of the CO2 plume
seismically, we plot the differences of the recovered images for each
monitor survey with respect to the baseline in
Figure~\ref{fig:LSRTMdiff}. These difference are computed via
$\mathbf{x}_j-\mathbf{x}_1$ for $j=2,\cdots,n_v$ where $\mathbf{x}_j$
are the images recovered independently. From these images, we observe
that time-lapse signal is visible albeit it is weak while the overall
image is plagued by strong artifacts, making it difficult to monitor the
CO2 plume.

To quantify the degree of repeatability of the time-lapse images, we
follow \citet{kragh2002seismic} and use the normalized root mean square
(NRMS) values defined as
\begin{equation}
NRMS(\mathbf{x}_1,\mathbf{x}_j)=\frac{200\times RMS(\mathbf{x}_1-\mathbf{x}_j)}{RMS(\mathbf{x}_1)+RMS(\mathbf{x}_j)}, \quad j=2,\cdots,n_v.
\label{eq:NRMS}
\end{equation}
 By definition, NRMS values range from $0$ to $200$ in percentage where
a smaller value indicates recovery results are more repeatable, with
$10$ percent as acceptable by today's best 4D practices. These NRMS
values are calculated only using the regions without time-lapse
difference.

\vspace*{-0.15cm}

\subsection{Joint recovery}\label{joint-recovery}

\vspace*{-0.15cm}

\begin{figure}
\centering
\includegraphics[width=1.000\hsize]{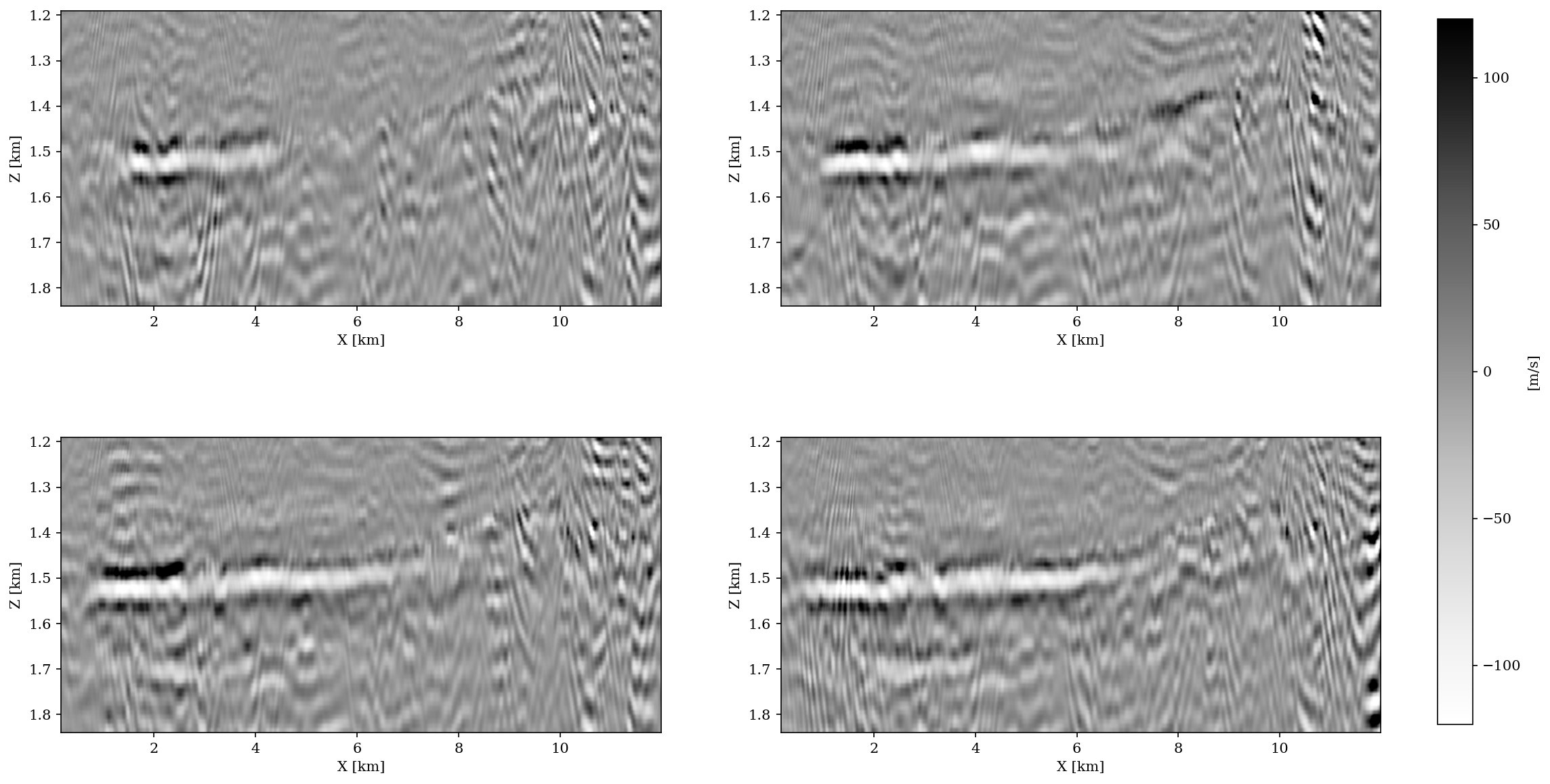}
\caption{Difference plots w.r.t. the baseline survey yielded by joint
recovery. The NRMS values computed for each pair are
$\{9.07\%, 8.83\%, 9.62\%, 11.23\%\}$.}\label{fig:JRMdiff}
\end{figure}

Finally, we recover the common component and innovations jointly via
Equation~\ref{eqjrm} with the same number of data passes (same number of
PDE solves) and the same random subsets of shots as during the
independent recovery. For each vintage, a different threshold
$\lambda_j$ is chosen based on the $90$th percentile of the
corresponding subset of $|\mathbf{u}_k|$ at the first iteration.
Figure~\ref{fig:JRMdiff} shows the difference plots. Thanks to the JRM,
we observe significantly fewer artifacts except for some artifacts in
the middle due to overlapping shots during the early iterations, a
problem that can easily be fixed. The improved quality of the difference
plots is also reflected in the NMRS values, which are now well within
the acceptable range. The scripts to reproduce the experiments in this
expanded abstract are made available on the
\href{https://slim.gatech.edu}{SLIM} GitHub page
\url{https://github.com/slimgroup/Software.SEG2021}.

\vspace*{-0.45cm}

\section{Discussion and conclusions}\label{discussion-and-conclusions}

\vspace*{-0.3cm}

Time-lapse seismic monitoring of carbon storage and sequestration is
difficult due to the fact that the velocity changes induced by the
growing CO2 plume are relatively small, making it challenging to detect
these changes seismically especially when the surveys are poorly sampled
and not replicated. By using the joint recovery model, we demonstrably
overcome this problem by producing significant improvements in the
difference plots associated with a realistic 60 year sequestration
scenario. While similar improvements by the joint recovery model have
been reported in the literature, the benefit of this approach has not
yet been demonstrated for multiple monitor surveys. Neither has this
model been used to monitor the development of a CO2 plume in a realistic
geological setting in the North Sea that is currently being considered
as a site for CO2 sequestration. While these initial results yielding
NMRS values of $10$\% on average are encouraging, many challenges remain
in the development of a low-cost 3D seismic monitoring system designed
to control and optimize the CO2 sequestration while minimizing its risk.

\vspace*{-0.45cm}

\section{Acknowledgement}\label{acknowledgement}

\vspace*{-0.3cm}

We would like to thank Charles Jones for the constructive discussion and
thank BG Group for providing the Compass model. The CCS project
information is taken from the Strategic UK CCS Storage Appraisal
Project, funded by DECC, commissioned by the ETI and delivered by Pale
Blue Dot Energy, Axis Well Technology and Costain. The information
contains copyright information licensed under
\href{https://s3-eu-west-1.amazonaws.com/assets.eti.co.uk/legacyUploads/2016/04/ETI-licence-v2.1.pdf}{ETI
Open Licence}. This research was carried out with the support of Georgia
Research Alliance and partners of the ML4Seismic Center.

\bibliography{yin2021SEGcts}

\begin{thebibliography}{38}
\providecommand{\natexlab}[1]{#1}
\providecommand{\url}[1]{\texttt{#1}}
\expandafter\ifx\csname urlstyle\endcsname\relax
  \providecommand{\doi}[1]{doi: #1}\else
  \providecommand{\doi}{doi: \begingroup \urlstyle{rm}\Url}\fi

\bibitem[Arts et~al.(2003)Arts, Eiken, Chadwick, Zweigel, {van der Meer}, and
  Zinszner]{ARTS2003347}
R.~Arts, O.~Eiken, A.~Chadwick, P.~Zweigel, L.~{van der Meer}, and B.~Zinszner.
\newblock Monitoring of co2 injected at sleipner using time lapse seismic data.
\newblock In J.~Gale and Y.~Kaya, editors, \emph{Greenhouse Gas Control
  Technologies - 6th International Conference}, pages 347--352. Pergamon,
  Oxford, 2003.
\newblock ISBN 978-0-08-044276-1.
\newblock \doi{https://doi.org/10.1016/B978-008044276-1/50056-8}.
\newblock URL
  \url{https://www.sciencedirect.com/science/article/pii/B9780080442761500568}.

\bibitem[Avseth et~al.(2010)Avseth, Mukerji, and Mavko]{avseth2010quantitative}
Per Avseth, Tapan Mukerji, and Gary Mavko.
\newblock \emph{Quantitative seismic interpretation: Applying rock physics
  tools to reduce interpretation risk}.
\newblock Cambridge university press, 2010.

\bibitem[Bharadwaj et~al.(2020)Bharadwaj, Li, and Demanet]{bharadwaj2020symae}
Pawan Bharadwaj, Matt Li, and Laurent Demanet.
\newblock Symae: an autoencoder with embedded physical symmetries for passive
  time-lapse monitoring.
\newblock In \emph{SEG Technical Program Expanded Abstracts 2020}, pages
  1586--1590. Society of Exploration Geophysicists, 2020.

\bibitem[Costa(2006)]{costa2006permeability}
Antonio Costa.
\newblock Permeability-porosity relationship: A reexamination of the
  kozeny-carman equation based on a fractal pore-space geometry assumption.
\newblock \emph{Geophysical research letters}, 33\penalty0 (2), 2006.

\bibitem[Furre et~al.(2017)Furre, Eiken, Alnes, Vevatne, and
  Kiær]{FURRE20173916}
Anne-Kari Furre, Ola Eiken, Håvard Alnes, Jonas~Nesland Vevatne, and
  Anders~Fredrik Kiær.
\newblock 20 years of monitoring co2-injection at sleipner.
\newblock \emph{Energy Procedia}, 114:\penalty0 3916--3926, 2017.
\newblock ISSN 1876-6102.
\newblock \doi{https://doi.org/10.1016/j.egypro.2017.03.1523}.
\newblock URL
  \url{https://www.sciencedirect.com/science/article/pii/S1876610217317174}.
\newblock 13th International Conference on Greenhouse Gas Control Technologies,
  GHGT-13, 14-18 November 2016, Lausanne, Switzerland.

\bibitem[Herrmann(2010)]{herrmann2010GEOPrsg}
Felix~J. Herrmann.
\newblock Randomized sampling and sparsity: getting more information from fewer
  samples.
\newblock \emph{Geophysics}, 75\penalty0 (6):\penalty0 WB173--WB187, 12 2010.
\newblock \doi{10.1190/1.3506147}.
\newblock URL
  \url{https://slim.gatech.edu/Publications/Public/Journals/Geophysics/2010/herrmann2010GEOPrsg/herrmann2010GEOPrsg.pdf}.

\bibitem[Herrmann and Hennenfent(2008)]{herrmann2008GJInps}
Felix~J. Herrmann and Gilles Hennenfent.
\newblock Non-parametric seismic data recovery with curvelet frames.
\newblock \emph{Geophysical Journal International}, 173:\penalty0 233--248, 04
  2008.
\newblock \doi{10.1111/j.1365-246X.2007.03698.x}.
\newblock URL
  \url{https://slim.gatech.edu/Publications/Public/Journals/GeophysicalJournalInternational/2008/herrmann2008GJInps.pdf}.

\bibitem[Janiszewski et~al.(2014)Janiszewski, Brewer, and
  Mosher]{janiszewski2014improvements}
F~Janiszewski, J~Brewer, and C~Mosher.
\newblock Improvements in the efficiency of ocean bottom sensor surveys through
  the use of multiple independent seismic sources.
\newblock In \emph{EAGE Workshop on Land and Ocean Bottom-Broadband Full
  Azimuth Seismic Surveys}, volume 2014, pages 1--3. European Association of
  Geoscientists \& Engineers, 2014.

\bibitem[Kaur et~al.(2020)Kaur, Sun, Zhong, and Fomel]{kaur2020time}
Harpreet Kaur, Alexander Sun, Zhi Zhong, and Sergey Fomel.
\newblock Time-lapse seismic data inversion for estimating reservoir parameters
  using deep learning.
\newblock In \emph{SEG Technical Program Expanded Abstracts 2020}, pages
  1720--1724. Society of Exploration Geophysicists, 2020.

\bibitem[Kragh and Christie(2002)]{kragh2002seismic}
ED~Kragh and Phil Christie.
\newblock Seismic repeatability, normalized rms, and predictability.
\newblock \emph{The Leading Edge}, 21\penalty0 (7):\penalty0 640--647, 2002.

\bibitem[Li et~al.(2013)Li, Mosher, Morley, Ji, and Brewer]{li2013joint}
Chengbo Li, Charles~C Mosher, Larry~C Morley, Yongchang Ji, and Joel~D Brewer.
\newblock Joint source deblending and reconstruction for seismic data.
\newblock In \emph{SEG Technical Program Expanded Abstracts 2013}, pages
  82--87. Society of Exploration Geophysicists, 2013.

\bibitem[Li et~al.(2020)Li, Xu, Harris, and Darve]{li2020coupled}
Dongzhuo Li, Kailai Xu, Jerry~M Harris, and Eric Darve.
\newblock Coupled time-lapse full-waveform inversion for subsurface flow
  problems using intrusive automatic differentiation.
\newblock \emph{Water Resources Research}, 56\penalty0 (8):\penalty0
  e2019WR027032, 2020.

\bibitem[Li et~al.(2012)Li, Aravkin, van Leeuwen, and Herrmann]{Li11TRfrfwi}
Xiang Li, Aleksandr~Y. Aravkin, Tristan van Leeuwen, and Felix~J. Herrmann.
\newblock Fast randomized full-waveform inversion with compressive sensing.
\newblock \emph{Geophysics}, 77\penalty0 (3):\penalty0 A13--A17, 05 2012.
\newblock \doi{10.1190/geo2011-0410.1}.
\newblock URL
  \url{https://slim.gatech.edu/Publications/Public/Journals/Geophysics/2012/Li11TRfrfwi/Li11TRfrfwi.pdf}.

\bibitem[Li(2015)]{li2015weighted}
Xiaowei Li.
\newblock \emph{A weighted $\ell_1$-minimization for distributed compressive
  sensing}.
\newblock PhD thesis, University of British Columbia, 2015.

\bibitem[Lorenz et~al.(2014)Lorenz, Schopfer, and Wenger]{lorenz2014linearized}
Dirk~A Lorenz, Frank Schopfer, and Stephan Wenger.
\newblock The linearized bregman method via split feasibility problems:
  Analysis and generalizations.
\newblock \emph{SIAM Journal on Imaging Sciences}, 7\penalty0 (2):\penalty0
  1237--1262, 2014.

\bibitem[Louboutin and Herrmann(2021)]{louboutin2021ultra}
Mathias Louboutin and Felix~J Herrmann.
\newblock Ultra-low memory seismic inversion with randomized trace estimation.
\newblock \emph{arXiv preprint arXiv:2104.00794}, 2021.

\bibitem[Louboutin et~al.(2019)Louboutin, Lange, Luporini, Kukreja, Witte,
  Herrmann, Velesko, and Gorman]{louboutin2018dae}
Mathias Louboutin, Michael Lange, Fabio Luporini, Navjot Kukreja, Philipp~A.
  Witte, Felix~J. Herrmann, Paulius Velesko, and Gerard~J. Gorman.
\newblock Devito (v3.1.0): an embedded domain-specific language for finite
  differences and geophysical exploration.
\newblock \emph{Geoscientific Model Development}, 2019.
\newblock \doi{10.5194/gmd-12-1165-2019}.
\newblock URL
  \url{https://slim.gatech.edu/Publications/Public/Journals/GMD/2019/louboutin2018dae/louboutin2018dae.pdf}.
\newblock (Geoscientific Model Development).

\bibitem[Lumley et~al.(1997)Lumley, Behrens, and Wang]{lumley1997assessing}
David~E Lumley, Ronald~A Behrens, and Zhijing Wang.
\newblock Assessing the technical risk of a 4d seismic project.
\newblock In \emph{SEG Technical Program Expanded Abstracts 1997}, pages
  894--897. Society of Exploration Geophysicists, 1997.

\bibitem[Luporini et~al.(2020)Luporini, Lange, Louboutin, Kukreja, Huckelheim,
  Yount, Witte, Kelly, Gorman, and Herrmann]{luporini2018aap}
Fabio Luporini, Michael Lange, Mathias Louboutin, Navjot Kukreja, Jan
  Huckelheim, Charles Yount, Philipp~A. Witte, Paul H.~J. Kelly, Gerard~J.
  Gorman, and Felix~J. Herrmann.
\newblock Architecture and performance of devito, a system for automated
  stencil computation.
\newblock \emph{ACM Trans. Math. Softw.}, 46\penalty0 (1), 04 2020.
\newblock \doi{10.1145/3374916}.
\newblock URL
  \url{https://slim.gatech.edu/Publications/Public/Journals/ACMTOMS/2020/luporini2018aap/luporini2018aap.pdf}.
\newblock (ACM Trans. Math. Softw.).

\bibitem[Maharramov et~al.(2019)Maharramov, Willemsen, Routh, Peacock,
  Froneberger, Robinson, Bear, and Lazaratos]{maharramov2019integrated}
Musa Maharramov, Bram Willemsen, Partha~S Routh, Emily~F Peacock, Mark
  Froneberger, Alana~P Robinson, Glenn~W Bear, and Spyros~K Lazaratos.
\newblock Integrated kinematic time-lapse inversion workflow leveraging
  full-waveform inversion and machine learning.
\newblock \emph{The Leading Edge}, 38\penalty0 (12):\penalty0 943--948, 2019.

\bibitem[Oghenekohwo and Herrmann(2015)]{oghenekohwo2015CSEGctl}
Felix Oghenekohwo and Felix~J. Herrmann.
\newblock Compressive time-lapse seismic data processing using shared
  information.
\newblock In \emph{CSEG Annual Conference Proceedings}, 05 2015.
\newblock URL
  \url{https://slim.gatech.edu/Publications/Public/Conferences/CSEG/2015/oghenekohwo2015CSEGctl/oghenekohwo2015CSEGctl.pdf}.
\newblock (CSEG, Calgary).

\bibitem[Oghenekohwo and Herrmann(2017)]{oghenekohwo2017EAGEitl}
Felix Oghenekohwo and Felix~J. Herrmann.
\newblock Improved time-lapse data repeatability with randomized sampling and
  distributed compressive sensing.
\newblock In \emph{EAGE Annual Conference Proceedings}, 06 2017.
\newblock \doi{10.3997/2214-4609.201701389}.
\newblock URL
  \url{https://slim.gatech.edu/Publications/Public/Conferences/EAGE/2017/oghenekohwo2017EAGEitl/oghenekohwo2017EAGEitl.html}.
\newblock (EAGE, Paris).

\bibitem[Oghenekohwo et~al.(2017)Oghenekohwo, Wason, Esser, and
  Herrmann]{oghenekohwo2016GEOPctl}
Felix Oghenekohwo, Haneet Wason, Ernie Esser, and Felix~J. Herrmann.
\newblock Low-cost time-lapse seismic with distributed compressive
  sensing{\textendash}-part 1: exploiting common information among the
  vintages.
\newblock \emph{Geophysics}, 82\penalty0 (3):\penalty0 P1--P13, 05 2017.
\newblock \doi{10.1190/geo2016-0076.1}.
\newblock URL
  \url{https://slim.gatech.edu/Publications/Public/Journals/Geophysics/2017/oghenekohwo2016GEOPctl/oghenekohwo2016GEOPctl.html}.
\newblock (Geophysics).

\bibitem[Qu and Verschuur(2017)]{qu2017simultaneous}
Shan Qu and Dirk Verschuur.
\newblock Simultaneous joint migration inversion for semicontinuous time-lapse
  seismic data.
\newblock In \emph{SEG Technical Program Expanded Abstracts 2017}, pages
  5808--5813. Society of Exploration Geophysicists, 2017.

\bibitem[Quei{\ss}er and Singh(2013)]{queisser2013full}
Manuel Quei{\ss}er and Satish~C Singh.
\newblock Full waveform inversion in the time lapse mode applied to co2 storage
  at sleipner.
\newblock \emph{Geophysical prospecting}, 61\penalty0 (3):\penalty0 537--555,
  2013.

\bibitem[Thomsen(1986)]{thomsen1986weak}
Leon Thomsen.
\newblock Weak elastic anisotropy.
\newblock \emph{Geophysics}, 51\penalty0 (10):\penalty0 1954--1966, 1986.

\bibitem[Tian et~al.(2018)Tian, Wei, Li, Oppert, and Hennenfent]{tian2018joint}
Yue Tian, Lei Wei, Chang Li, Shauna Oppert, and Gilles Hennenfent.
\newblock Joint sparsity recovery for noise attenuation.
\newblock In \emph{SEG Technical Program Expanded Abstracts 2018}, pages
  4186--4190. Society of Exploration Geophysicists, 2018.

\bibitem[Wason et~al.(2015)Wason, Oghenekohwo, and Herrmann]{wason2015EAGEcsm}
Haneet Wason, Felix Oghenekohwo, and Felix~J. Herrmann.
\newblock Compressed sensing in {4-D marine}---recovery of dense time-lapse
  data from subsampled data without repetition.
\newblock In \emph{EAGE Annual Conference Proceedings}, 06 2015.
\newblock \doi{10.3997/2214-4609.201413088}.
\newblock URL
  \url{https://slim.gatech.edu/Publications/Public/Conferences/EAGE/2015/wason2015EAGEcsm/wason2015EAGEcsm.html}.
\newblock (EAGE, Madrid).

\bibitem[Wason et~al.(2017)Wason, Oghenekohwo, and Herrmann]{wason2016GEOPctl}
Haneet Wason, Felix Oghenekohwo, and Felix~J. Herrmann.
\newblock Low-cost time-lapse seismic with distributed compressive
  sensing{\textendash}-part 2: impact on repeatability.
\newblock \emph{Geophysics}, 82\penalty0 (3):\penalty0 P15--P30, 05 2017.
\newblock \doi{10.1190/geo2016-0252.1}.
\newblock URL
  \url{https://slim.gatech.edu/Publications/Public/Journals/Geophysics/2017/wason2016GEOPctl/wason2016GEOPctl.html}.
\newblock (Geophysics).

\bibitem[Wei et~al.(2018)Wei, Tian, Li, Oppert, and Hennenfent]{wei2018improve}
Lei Wei, Yue Tian, Chang Li, Shauna Oppert, and Gilles Hennenfent.
\newblock Improve 4d seismic interpretability with joint sparsity recovery.
\newblock In \emph{SEG Technical Program Expanded Abstracts 2018}, pages
  5338--5342. Society of Exploration Geophysicists, 2018.

\bibitem[Witte et~al.(2019{\natexlab{a}})Witte, Louboutin, Kukreja, Luporini,
  Lange, Gorman, and Herrmann]{witte2018alf}
Philipp~A. Witte, Mathias Louboutin, Navjot Kukreja, Fabio Luporini, Michael
  Lange, Gerard~J. Gorman, and Felix~J. Herrmann.
\newblock A large-scale framework for symbolic implementations of seismic
  inversion algorithms in julia.
\newblock \emph{Geophysics}, 84\penalty0 (3):\penalty0 F57--F71, 03
  2019{\natexlab{a}}.
\newblock \doi{10.1190/geo2018-0174.1}.
\newblock URL
  \url{https://slim.gatech.edu/Publications/Public/Journals/Geophysics/2019/witte2018alf/witte2018alf.pdf}.
\newblock (Geophysics).

\bibitem[Witte et~al.(2019{\natexlab{b}})Witte, Louboutin, Luporini, Gorman,
  and Herrmann]{witte2018cls}
Philipp~A. Witte, Mathias Louboutin, Fabio Luporini, Gerard~J. Gorman, and
  Felix~J. Herrmann.
\newblock Compressive least-squares migration with on-the-fly fourier
  transforms.
\newblock \emph{Geophysics}, 84\penalty0 (5):\penalty0 R655--R672, 08
  2019{\natexlab{b}}.
\newblock \doi{10.1190/geo2018-0490.1}.
\newblock URL
  \url{https://slim.gatech.edu/Publications/Public/Journals/Geophysics/2019/witte2018cls/witte2018cls.pdf}.
\newblock (Geophysics).

\bibitem[Xuan et~al.(2020)Xuan, Malik, Zhang, Guo, and
  Huang]{xuan2020deblending}
Yi~Xuan, Raheel Malik, Zhigang Zhang, Manhong Guo, and Yi~Huang.
\newblock Deblending of obn ultralong-offset simultaneous source acquisition.
\newblock In \emph{SEG Technical Program Expanded Abstracts 2020}, pages
  2764--2768. Society of Exploration Geophysicists, 2020.

\bibitem[Yang et~al.(2015)Yang, Meadows, Inderwiesen, Landa, Malcolm, and
  Fehler]{yang2015double}
Di~Yang, Mark Meadows, Phil Inderwiesen, Jorge Landa, Alison Malcolm, and
  Michael Fehler.
\newblock Double-difference waveform inversion: Feasibility and robustness
  study with pressure data.
\newblock \emph{Geophysics}, 80\penalty0 (6):\penalty0 M129--M141, 2015.

\bibitem[Yang et~al.(2016)Yang, Liu, Morton, Malcolm, and Fehler]{yang2016time}
Di~Yang, Faqi Liu, Scott Morton, Alison Malcolm, and Michael Fehler.
\newblock Time-lapse full-waveform inversion with ocean-bottom-cable data:
  Application on valhall field.
\newblock \emph{Geophysics}, 81\penalty0 (4):\penalty0 R225--R235, 2016.

\bibitem[Yang et~al.(2020)Yang, Fang, Witte, and Herrmann]{yang2020tdsp}
Mengmeng Yang, Zhilong Fang, Philipp~A. Witte, and Felix~J. Herrmann.
\newblock Time-domain sparsity promoting least-squares reverse time migration
  with source estimation.
\newblock \emph{Geophysical Prospecting}, 68\penalty0 (9):\penalty0 2697--2711,
  08 2020.
\newblock \doi{10.1111/1365-2478.13021}.
\newblock URL
  \url{https://slim.gatech.edu/Publications/Public/Journals/GeophysicalProspecting/2020/yang2020tdsp/yang2020tdsp.html}.
\newblock (Geophysical Prospecting).

\bibitem[Yin et~al.(2008)Yin, Osher, Goldfarb, and Darbon]{yin2008bregman}
Wotao Yin, Stanley Osher, Donald Goldfarb, and Jerome Darbon.
\newblock Bregman iterative algorithms for $\backslash$ell\_1-minimization with
  applications to compressed sensing.
\newblock \emph{SIAM Journal on Imaging sciences}, 1\penalty0 (1):\penalty0
  143--168, 2008.

\bibitem[Zhang and Huang(2013)]{zhang2013double}
Zhigang Zhang and Lianjie Huang.
\newblock Double-difference elastic-waveform inversion with prior information
  for time-lapse monitoring.
\newblock \emph{Geophysics}, 78\penalty0 (6):\penalty0 R259--R273, 2013.

\end{thebibliography}

\end{document}